\documentstyle[preprint,aps]{revtex}
\begin{document}
%%%%%%%%%%%%%%%%%%%%%%%%%%%%%%%%%%%%%%%%%%%%%
\preprint{UCD--95--16}
\title{Exact and Approximate
Radiation Amplitude Zeros \\
--- Phenomenological Aspects}

\author{Tao Han}
\address{Davis Institute for High Energy Physics\\
Department of Physics, University of California, Davis 95616}

\maketitle

\begin{abstract}

We review the phenomenological  aspects  of the exact
and approximate Radiation Amplitude Zeros (RAZ) and discuss
the prospects of searches for these zeros at current
and future collider experiments.

\end{abstract}

\section*{I. \, Exact and Approximate Radiation Amplitude Zeros}

More than 15 years ago, the pioneer studies on vector-boson
pair production  \cite{brown1,brown2,mikaelian} revealed
a surprise: the angular distribution for $f_1 \bar f_2 \rightarrow W^- \gamma$
develops a pronounced zero \cite{mikaelian,brown2} at
\begin{equation}
{\rm cos}\theta=(Q_{f_1}+Q_{f_2})/(Q_{f_1} - Q_{f_2}) ,
\label{EQ:zero}
\end{equation}
where $\theta$ is the $W^-$ scattering angle with respect to the incident
fermion ($f_1$) direction in the center of mass (c.m.) frame, and
$Q_{f_i}$ the electric charge of fermion $f_i$.
Figure~\ref{wgzero} demonstrates this unusual angular distribution for
$e^- \nu, \,  d\bar u \rightarrow W^- \gamma$ processes, in which the
zero occurs at $\cos\theta=1, -1/3$, respectively.
The authors  of Ref. \cite{mikaelian}  stated in the abstract  that
``... We can offer no explanation for this behavior''.

In fact, it is not difficult to see what is happening for some
simple cases. Take $d\bar u \rightarrow W^- \gamma$
as an example.  There are three Feynman diagrams to contribute
at the Born level: a $t$-channel diagram
with an  amplitude proportional to $Q_u/t$, a $u$-channel diagram
proportional to $Q_d/u$, and an $s$-channel diagram
proportional to $Q_{W^-}/(s-M^2_W)$, where
 $t=(p_d^{} - p_W^{})^2=-{1 \over 2}(s-M^2_W)(1-\cos\theta)$.
Notice the  charge relation in the Standard Model  $Q_d - Q_u=Q_{W^-}$,
and the kinematical relation $s-M^2_W=-t-u$, one can easily cast the
amplitude  into the form
\begin{equation}
{\cal M} \sim ( {Q_u \over t} + {Q_d \over u}) F(\sigma_i, \lambda_i, p_i),
\label{factorwg}
\end{equation}
where $F(\sigma_i, \lambda_i, p_i)$ denotes a reduced matrix element as
a function of the  fermion helicity
$\sigma_i$, vector-boson polarization $\lambda_i$ and the
external momenta $p_i$.
We see immediately that this amplitude develops a zero at a special
angle determined by Eq.  \ref{EQ:zero}.

Not long after this discovery, several groups \cite{zhuetal,brodsky,criteria}
further examined this interesting feature. It was found that
in gauge theories, any tree-level 4-particle  (spin $\le 1$)
Feynman amplitudes  with  one or more massless gauge particles
can be factorized into
two factors, one of which contains the dependence of internal
quantum numbers (such as charges) and the other contains the
dependence of spin and polarization indices \cite{zhuetal}.
This factorization is a special case for a more general theorem \cite{brodsky},
which states that {\it for a tree-level  $n$-particle
(spin $\le 1$) amplitude with one photon,
the amplitude develops a zero when the factor $Q_i/p_i^{} \cdot q$  are
equal for $i=1,2...n-1$, where $Q_i$ and $p_i^{}$ are the charge and momentum
for the $i^{th}$ particle, respectively, and $q$ the photon momentum.}
There is certainly a deeper explanation for this phenomena, having
something to do with the relationship between the internal gauge symmetry
and the space-time symmetry. One can
find a very nice discussion in Bob Brown's talk  at this conference \cite{bob},
or from the classical papers on this subject  \cite{brodsky}.
Following the literature,   we will call  those zeros
Radiation Amplitude Zeros (RAZ)  \cite{general}.

It should be noted,  however,  that
\begin{itemize}
  \item  not all of the RAZ
occur in  physical region --- in fact, most of  them
do not.  The above theorem can be  translated into  an intuitive
necessary condition for RAZ to occur in physical region:
{\it along with a massless gauge boson, the other
particles involved in the process must have the same sign of electric charges}.
We will call this condition the ``same-sign rule''.
  \item although loop diagrams (and bubbles) do not significantly alter the
nature of RAZ \cite{loops},  higher order real emissions spoil the RAZ
\cite{realg,bhowg,ohnemus}. It was suggested \cite{bhowg}
that one can regain the
Born-level kinematics by vetoing additional final state particles, thus
recovering an ``approximate'' zero in practice.

\end{itemize}

It is natural to ask what may happen in  a theory
with  a spontaneously  broken gauge symmetry, such as
the Standard Model (SM). It is conceivable that  the radiation of a
$Z$-boson may have some similarity to that of a photon.  For the case
of   $d\bar u \rightarrow W^- Z$, the amplitude can be written as \cite{wzzero}
\begin{equation}
{\cal M} \sim  X F_X^{}(\sigma_i, \lambda_i, p_i)  +
Y F_Y^{}(\sigma_i, \lambda_i, p_i),
\label{factorwz}
\end{equation}
where $X$ and $Y$ are combinations of coupling factors
\begin{eqnarray}
X = {s \over 2} \, \left(  {g_{-}^{f_1} \over u}
                         + {g_{-}^{f_2} \over t} \right) \, , \ \
Y = g_{-}^{f_1} \, {M_Z^2 \,s\, \over 2\,u\, (s - M_W^2) } \>,
%\nonumber
\end{eqnarray}
with the left-handed neutral current couplings
$g^{f_1}_{-} - g^{f_2}_{-} = Q_W \, \cot\theta_{\rm w} $, and
$F_{X,Y}^{}(\sigma_i, \lambda_i, p_i)$
contain the spin dependent part and is roughly proportional
to the product of the vector-boson wave functions
$\epsilon^*_{\rm w} \cdot \epsilon^*_{\rm z}$.
It is obvious that without  the $Y$-term, the
helicity amplitudes would factorize. In this case, all amplitudes would
simultaneously vanish for $g_-^{f_1}/u + g_-^{f_2}/t =0$,
analogous to the $W\gamma$ case in Eq.  \ref{factorwg}.
Since  $Y$ is directly proportional to $M_Z^2$, one may
naively expect   full factorization when $M_Z^2 \ll s$. In fact,
in the high energy limit, only three helicity amplitudes remain non-zero:
\begin{eqnarray}
{\cal M}(\lambda_{\rm w}=\pm,\lambda_{\rm z}=\mp)
&\longrightarrow&  {1 \over \sin\theta }\,
(\lambda_{\rm w} - \cos\theta)\,
\Bigl[  (g^{f_1}_{-} - g^{f_2}_{-} ) \cos\theta   - (g^{f_1}_{-} + g^{f_2}_{-}
) \Bigr] \>,
\nonumber \\
\noalign{\vskip 10pt}
{\cal M}(\lambda_{\rm w}=0,\lambda_{\rm z}=0)
&\longrightarrow&  {1 \over 2} \, \sin\theta \,
{M_Z \over M_W} \, (g^{f_2}_{-} -  g^{f_1}_{-}) \>.
%\nonumber
\end{eqnarray}
While  the dominant amplitudes ${\cal M}(\pm,\mp)$
fully factorize in the high energy limit,  ${\cal M}(0,0)$ behaves differently.
This can be traced to the special energy-dependence of the
polarization vectors for longitudinal vector bosons,
$\epsilon_{\rm v} \sim \sqrt{s}/M_V^{}$. Since the $Y$-term in
Eq.~\ref{factorwz}  goes like
$(M_Z^2/s)\, \epsilon^*_{\rm w} \cdot \epsilon^*_{\rm z}$,
the ${\cal M}(0,0)$ amplitude remains finite  at high energies.

The combined effect of the zero in ${\cal M}(\pm,\mp)$ and the relatively
small contributions from the remaining helicity amplitudes
results in an approximate zero for the $f_1\bar f_2\rightarrow W^\pm Z$
differential cross section at
\[   \cos\theta \simeq  (g^{f_1}_-+g^{f_2}_-)/(g^{f_1}_- - g^{f_2}_-)
 \simeq \left\{
\begin{array}{ll}
\phantom{+}{1\over 3}\tan^2\theta_{\rm w} \simeq 0.1
& \mbox{for $d \bar u \rightarrow W^{-} Z\>,$} \\
 - \tan^2\theta_{\rm w}  \simeq -0.3
& \mbox{for $e^{-} \bar \nu_e \rightarrow W^{-} Z\>.$}
\end{array}
\right. \]
This is illustrated in Fig.~\ref{wzzerofig} where  the
differential cross sections  are shown for
$e^-\bar\nu_e \rightarrow W^-Z$ and $d\bar u\rightarrow W^-Z$
for $(\lambda_{\rm w},\lambda_{\rm z})=(\pm,\mp)$ and $(0,0)$,
as well as the unpolarized cross
section, which is obtained by summing over all $W$- and $Z$-boson helicity
combinations (solid line).
For both reactions, the total differential cross section displays a
pronounced minimum at
the location of the zero in ${\cal M}(\pm,\mp)$. Due to the
$1/\sin\theta$ behaviour of ${\cal M}(\pm,\mp)$, the $(+,-)$ and $(-,+)$
amplitudes dominate outside of the region of the zero. In order to
demonstrate the influence of the zero in ${\cal M}(\pm,\mp)$ on the
total angular differential cross section,  the
$\cos\theta$ distribution for $e^+e^-\rightarrow ZZ$
has been included in  Fig.~\ref{wzzerofig}a).  The zero in the $(\pm,\mp)$
amplitudes causes the minimum in the $WZ$ case to be much more
pronounced than the minimum in $e^+e^-\rightarrow ZZ$.

It is important to note that the RAZ are the direct results from subtle
gauge cancellation.  Non-standard couplings, such as those
$\Delta g_1$, $\Delta\kappa$ and $\lambda$ \cite{hagietal}
spoil these cancellations and eliminate the (approximate) zeros.
This can be seen from  the additional contributions to the SM
amplitudes, for the $W\gamma$ process,
\begin{eqnarray}
%\Delta{\cal M}_{{\rm w}\gamma}(\pm,\mp) &=& 0\>,
%\label{EQ:PLUSMIN} \\
%
%\noalign{\vskip 8pt}
%
\Delta{\cal M}_{{\rm w}\gamma}(\pm,\pm) &=& {F\over 2} \>  \sin\theta \,
\biggl[ \Delta\kappa + {\lambda \over r_{\rm w} } \biggr] \>,\\
\noalign{\vskip 8pt}
\Delta{\cal M}_{{\rm w}\gamma}(0,\pm) &=& {F\over 2} \>
{ (1 + \lambda_\gamma \cos\theta ) \over \sqrt{2 r_{\rm w}} } \,
\Bigl[ \Delta\kappa + \lambda \Bigr] \>,
\end{eqnarray}
where $F= V_{f_1 f_2} e^2/\sqrt 2 \sin\theta_{\rm w}$ and $r_{\rm v}=M^2_V/s$;
the corresponding contributions to the $WZ$ production amplitudes are
\begin{eqnarray}
%\Delta{\cal M}_{\rm wz}(\pm,\mp) &=& 0\>,
%\label{EQ:PLUSMIN} \\
%
%\noalign{\vskip 8pt}
%
\Delta{\cal M}_{\rm wz}(\pm,\pm) &=& {F\over 2} \>
{Q_W \, \cot\theta_{\rm w} \over 1 - r_{\rm w} } \> \beta \sin\theta \,
\biggl[ \Delta g_1 + \Delta\kappa + {\lambda \over r_{\rm w} } \biggr] \>,\\
\noalign{\vskip 8pt}
\Delta{\cal M}_{\rm wz}(0,0) &=& {F\over 2} \>
{Q_W \, \cot\theta_{\rm w} \over 1 - r_{\rm w} } \>
{\beta \sin\theta \over \sqrt{2 r_{\rm w} 2 r_{\rm z}} } \, 2 \,
\Bigl[ \Delta g_1 (1 + r_{\rm w}) + \Delta\kappa\, r_{\rm z} \Bigr] \>,
\label{EQ:ZEROZERO} \\
\noalign{\vskip 8pt}
\Delta{\cal M}_{\rm wz}(\pm,0) &=& {F\over 2} \>
{Q_W \, \cot\theta_{\rm w} \over 1 - r_{\rm w} } \>
{\beta (1- \lambda_{\rm w}\cos\theta ) \over \sqrt{2 r_{\rm z}} } \,
\biggl[ 2\, \Delta g_1 + \lambda \, {r_{\rm z}\over r_{\rm w}} \biggr] \>,\\
\noalign{\vskip 8pt}
\Delta{\cal M}_{\rm wz}(0,\pm) &=& {F\over 2} \>
{Q_W \, \cot\theta_{\rm w} \over 1 - r_{\rm w} } \>
{\beta (1 + \lambda_{\rm z}\cos\theta ) \over \sqrt{2 r_{\rm w}} } \,
\Bigl[ \Delta g_1 + \Delta\kappa + \lambda \Bigr] \>,
\label{EQ:NSMHELAMP}
\end{eqnarray}
where $\beta=[(1- r_{\rm w}- r_{\rm z})^2 - 4 r_{\rm w} r_{\rm z}]^{1/2}$.
Due to angular momentum  conservation, the $(\pm,\mp)$ amplitudes which
dominate in the SM do not receive any contributions from the
anomalous couplings. The amplitude zeros  in these two
helicity configurations for both
$W\gamma$ and $WZ$ channels thus remain exact.
All other helicity amplitudes are
modified in the presence of non-standard $WW\gamma$/$WWZ$ couplings.
At high energies the anomalous contributions grow proportional to
$\sqrt{s}$ ($s$) for $\Delta\kappa$ ($\Delta g_1$ and $\lambda$)
and eventually dominate the cross section. The nature of the
RAZ is thus sensitive to new physics in the vector-boson sector.

\section*{II. \, Prospects of Experimental Searches for RAZ}

Clearly, the radiation amplitude zeros (RAZ) are a very interesting
feature of gauge theories and  it would be desirable to experimentally
observe this distinctive phenomena. However,  we emphasize that
studying these ``zeros'' is not to search for ``nothing''. Rather,
we would hope to find new physics in the vector-boson sector
and the amplitudes near RAZ are especially sensitive to
the deviation from the SM.
This is the motivation to examine the feasibility of experimental
searches for RAZ.

\subsection*{A. \, $W\gamma$ Production at Hadron Colliders:  \,
$p \bar p, pp \rightarrow W^\pm \gamma  \rightarrow l^\pm \nu \gamma$}

The successful $p \bar p$ collider experiments at the Fermilab
Tevatron may provide suitable environment for searching for
RAZ and for testing the anomalous gauge boson couplings  \cite{talks}.
However, it is non-trivial to carry out the searches for the RAZ
experimentally.  The problems, both theoretical and experimental,
include:

\begin{enumerate}
 \item   {\it reconstruction of the $q \bar q $ c.m. frame:}
it is impossible to non-ambiguously reconstruct  the parton c.m. frame
to define the scattering angle to obtain $d\sigma/d\cos\theta$ since the
reconstruction  of  the neutrino momentum  ($p_\nu$) from  constraint
$(p^{}_l + p^{}_\nu)^2=M_W^2$
is subject to a two-fold  ambiguity \cite{pnu,hagi1}.
 \item    {\it  $z$-axis along the incident fermion moving direction:}  in
hadron
colliders, there are two types of  parton-level contributions to the
same final state:  $d_1 \bar u_2 \rightarrow W^- \gamma$ and
$\bar u_1 d_2 \rightarrow W^- \gamma$.  Since the polar angle  $\theta$
is defined with respect to incident fermion moving direction $\vec p^{}_d$,
it is then impossible to non-ambiguously identify the  direction of  $z$-axis
(along the $d$-quark).
In $p\bar p$ collisions at  Tevatron energies, due to the valence
quark dominance, the contribution from $d_1 \bar u_2$ is much larger
than that from $\bar u_1 d_2$, so that one can simply assign the $z$-axis
along the proton direction. However, in $pp$ collisions, those
contributions are equal, making the $z$-axis identification intrinsically
impossible.
\item {\it higher order corrections:} the RAZ in
$d \bar u \rightarrow W^- \gamma$ is exact  only for the $2\rightarrow 2$
Born-level process.  Additional jets  from higher order QCD radiation
\cite{realg,bhowg,ohnemus} will  spoil the subtle cancellation and thus fill up
the
zero.  One has to reject (or veto) the additional jets to recover the
Born-level kinematics \cite{bhowg}.
\item {\it  $W^-$ radiative decay:} for the  channel
$d \bar u \rightarrow W^- \gamma \rightarrow e^- \bar \nu_e \gamma$,
a single $W^-$ (Drell-Yan) production with subsequent radiative decay
$d \bar u \rightarrow W^- \rightarrow e^- \bar \nu_e \gamma$ gives
the same final state but different kinematical structure.  Those events
should be kept separated. This could be achieved by imposing a
transverse mass cut   \cite{hagi1,ulidieter,ulied} slightly above
$M^{}_W$,  $M_T(l^\pm \nu,\gamma) > 90$ GeV.
\item {\it  backgrounds:} the most severe background for the $W^- \gamma$
final state seems to be the misidentification of a photon from a jet
$j \rightarrow \gamma$, due to the much larger production
rate for $W^-j$. Good $\gamma$-$j$ discrimination factor is needed
to successfully identify the signal.
\end{enumerate}

The first attempt to realistically study the RAZ at the Tevatron was carried
out
in Ref. \cite{hagi1}.  Due to the two-fold
ambiguity in constructing the neutrino momentum,
the authors  studied two polar angle distributions $\cos \theta^*_+$
and $\cos \theta^*_-$, corresponding to the two solutions for
$\cos \theta^*_\nu > \cos \theta^*_e$ and
$\cos \theta^*_\nu < \cos \theta^*_e$, respectively.
Although one is unable to tell the correct $p_\nu$ solution on an
even-by-event basis,  it is seen from Fig.~\ref{hagifig} that
$\cos \theta^*_-$ reflects the zero location  better. This can be
understood in terms of the $V-A$ coupling. Namely, $e^-$ ($\bar \nu_e$)
prefers to move in the forward (backward)
direction so that  $\cos \theta^*_\nu < \cos \theta^*_e$, which corresponds
to the $\cos \theta^*_-$ solution. Fig.~\ref{hagifig} also demonstrates
the anomalous coupling effects  that tend to fill up the dip.

The RAZ in Eq.~\ref{EQ:zero} corresponds to  the photon rapidity
in c.m. frame
\begin{equation}
y^*_\gamma = {\frac{1}{2} }\ln
{\frac{1+\cos\theta_\gamma}{1-\cos\theta_\gamma}}
= {\frac{1}{2} }\ln \, (- {\frac{Q_2}{Q_1}}) ,
\label{EQ:zeroy}
\end{equation}
which gives $y^*_\gamma \simeq \pm 0.35$ for $W^\mp \gamma$ channel.
As a direct reflection of the RAZ, the photon rapidity
spectrum in the c.m. frame  develops a clear dip in the central
region after summing over the two solutions for $p^{}_\nu$
\cite{ulidieter,ulied}.
A problem arises when we include QCD radiative corrections \cite{ohnemus}.
Although moderate at the Tevatron energies,  the QCD corrections tend to
fill up the dip and  to increase the cross section in a similar way
as the anomalous couplings \cite{bhowg}.  Figure \ref{yqcd}
shows the differential cross section for the photon rapidity in the
reconstructed center of mass frame for the reaction
$p \bar p \to W^+ \gamma \to e^+ \nu_e \gamma$ at
$\protect\sqrt{s} = 1.8$~TeV in the SM.
The inclusive next-to-leading-order
(NLO) differential cross section [solid line in a)] is seen
to be significantly larger than the Born-level leading-order (LO)
approximation
[dot-dashed line in b)] and tend to fill in the dip near zero.  However,
the NLO $W\gamma +0$~jet exclusive differential cross section
(dotted line) is comparable to the Born-level LO result.
This important observation implies
that if  we study the 0-jet exclusive process
$p \bar p \rightarrow W \gamma$+  0-jet $\rightarrow e \nu_e$+ 0-jet,
namely, if we veto the extra jet(s) from higher order QCD processes,
we recover most of the feature in the Born level and thus regain the
sensitivity to study the anomalous couplings.

It is noted that the RAZ for
$p \bar p \rightarrow W^\pm \gamma \rightarrow e^\pm \nu_e$
occur in the central region $\cos\theta=\pm \frac{1}{3}$
(and  $y^*_{\gamma}\simeq 0$ averagely).
Therefore, deviations from the SM will largely happen in
high transverse momentum $p_T^{}(\gamma$) region.  This feature
has been carefully examined in a recent paper \cite{zgvswg}.
It is shown that, as a function of a cutoff on the photon transverse
momenta   $p_T^{min}(\gamma)$,  the ratio of  integrated cross sections
$R_{\gamma,l}=\sigma(\gamma Z)/\sigma(\gamma W)$  for
$Z\gamma$ process  (which has no  RAZ) and for
$W\gamma$  process (which has a RAZ)
is a clear indication of a zero behavior,
as shown in Fig.~\ref{zgwg}.  We see that in high $p_T^{}(\gamma)$
region, the rate for  $W\gamma$ process is significantly smaller
than that of  $Z\gamma$ process. In contrast, the ratio
versus a cutoff on a jet transverse momentum $p^{min}_T(j)$ for
$Zj$ and $Wj$ production is flat over a large $p^{}_T(j)$ range.
The advantage of looking at the cross section ratio versus
$p_T^{min}(\gamma)$ is to have avoided the c.m. frame  and
$z$-axis ambiguities,  while  the price to pay is to lose the information
about the exact  RAZ location.

Some more interesting observation has been made recently in Ref. \cite{ycorrl}.
Recall the rapidity in the lab frame ($y$) is a sum of  that in c.m. frame
$(y^* )$
and a term  reflecting the c.m. frame motion
\begin{equation}
y = y^* + \frac{1}{2} \ln \, (\frac{x_1}{x_2}) ,
\end{equation}
where $x_{1,2}$ are the parton momentum fractions.

If we take the rapidity difference between the photon
and the $W$, then the difference
is invariant under the longitudinal boost. Therefore, the rapidity
correlation between $W$-$\gamma$ in the c.m. frame is preserved
in the lab frame
\begin{equation}
\Delta y =  y_\gamma - y^{}_W  =  y^*_\gamma - y^*_W \simeq -0.4.
\end{equation}
We thus have a chance to avoid the frame ambiguity if we choose
the variable in such a clever way.
Fig.~\ref{2dcorrl} demonstrates
the rapidity correlation in the lab frame \cite{ycorrl}.
We see an impressive ``valley'' for the rapidity correlation,
given by $y_\gamma - y^{}_W  \simeq -0.4$.
In order to implement this idea more realistically,
we must use the final state momentum
of $l^\pm$, rather than that of $W$. Fortunately, based on helicity
arguments, the charged lepton in $W\gamma$ process goes
dominantly along the $W$ moving direction, so that $W$-$\gamma$
rapidity correlation is largely preserved,
\begin{equation}
\Delta\eta(\gamma,l) = \eta(\gamma) - \eta(l) \simeq -0.3  .
\end{equation}
One could  directly study
the rapidity difference, which would have the advantage
for higher statistics than the double differential cross section. This is shown
in Fig.~\ref{1dcorrl}.  The curve for the pseudorapidity difference in the SM
(solid) presents a clear dip at  -0.3.
The authors of  Ref.  \cite{ycorrl} have also estimated
the error bars  for expected statistical uncertainties with  an integrated
luminosity of  22~pb$^{-1}$. Since the CDF/D0 collaborations  have
accumulated about 100 pb$^{-1}$ each (at the time of writing),
one can anticipate that an experimental study
along this line may first  observe the clear dip reflecting the RAZ.
The effects from anomalous couplings are also demonstrated in the figure.

Finally, two remarks are in order. First,
we have thus far concentrated on Tevatron energies.
At the LHC, due to the more  severe problems  regarding  the
$z$-axis definition and much larger QCD
corrections to the Born amplitudes,  the conclusions in studying
the RAZ  seem rather pessimistic.  Secondly,  we have not discussed
much about the background issue. It turns out that if we could
achieve a $j\rightarrow \gamma$ misidentification factor at  a level
of 10$^{-3}$, the background  may not be too severe \cite{ycorrl,talks}.

\subsection*{B. \,  $W$ Radiative Decay: \,
$W^\pm \rightarrow l^\pm \nu \gamma$, \, $q \bar q' \gamma$}

It  was shown \cite{wdecay}
that the $W$-radiative decay, $W \rightarrow  f_1 \bar f_2 \gamma$,
also presents a RAZ. Refs. \cite{hagi1,ulidieter,ulied} studied the process
$p \bar p \rightarrow W^\pm \rightarrow e^\pm \nu \gamma$
at the Tevatron.  This process develops a zero at the
kinematical boundary $\cos\theta_{l\gamma} = -1$ in $W$-rest frame.
It can be effectively separated from the
$W\gamma$ associated production by imposing a transverse mass cut,
$M_T^{}(l^\pm \nu,\gamma)<90$ GeV;  and it also
has  larger statistics.  However, the RAZ is less pronounced
due to the single-zero behavior \cite{hagi1,wdecay} and
the difficulty for $W$-rest frame
reconstruction.  It is therefore less sensitive to anomalous couplings.

Ref.  \cite{wdecay2} discussed the zero in the hadronic
decay process $W^- \rightarrow  d \bar u \gamma$. In this case,
it is a double-zero as usual and the $W$-rest frame reconstruction may be
relatively easier. However, the event identification may be
difficult in hadron collider experiments; and it will suffer from low
statistics for $e^+e^- \rightarrow W^+ W^-$  with a radiative $W$
decay.

\subsection*{C. \, $WZ$ Production: \,
$p \bar p, pp \rightarrow W^\pm Z \rightarrow l^\pm \nu l^+l^-$}

As discussed earlier,  the Born amplitude for
$q_1 \bar q_2 \rightarrow W^\pm Z$ develops a zero at high energies
\cite{wzzero} at
$\cos\theta \simeq  \pm {1\over 3}\tan^2\theta_{\rm W}\approx\pm 0.1$.
Following the proposal of studying the rapidity correlation analogous
to $\Delta\eta(\gamma,\ell)$  for $p\bar p\rightarrow W^+\gamma$
process in Fig. \ref{1dcorrl}, one can examine the rapidity correlation
\cite{bhowz} via
$\Delta y(Z,\ell_1)=y(Z)-y(l_1)$ where $l_1$ is the charged lepton
from $W$ decay. Figure \ref{wzcorrl} shows the differential cross section
$d\sigma/ \Delta y$.   There is a dip near 0.1 as predicted in the SM
(solid curve), although it will not be easy to convincingly establish the
effect
due to the less pronounced dip for this channel and limited
number of $W^\pm Z\to\ell_1^\pm\nu_1\ell_2^+\ell_2^-$ events expected
(see the estimated statistical error bars in the figure for 10 fb$^{-1}$
luminosity).
At the LHC energies, the zero is further washed out due to larger QCD
radiative corrections and the $z$-axis ambiguity.

It is amusing to note \cite{wzzero}  that $e\nu_e$ or $\mu\nu_\mu$ collisions
above the $WZ$ threshold would in principle provide a clean environment
for event reconstruction.
The location of the zero at $\cos\theta \approx \pm 0.3$ is ideal
for experimental studies of the $W^\pm Z$ final state, unlike the case for
$e^-\bar \nu_e \rightarrow W^-\gamma$ where the zero is
located at the kinematical boundary  ($\cos\theta=1$) resulting
a single-zero.

\subsection*{D. \, $qq' \rightarrow  qq' \gamma$ And $eq \rightarrow  eq
\gamma$}

Certain single photon radiation processes in quark scattering, such as
\begin{equation}
 uu \rightarrow  uu\gamma,  \quad  u \bar d \rightarrow u \bar d\gamma, \quad
 dd \rightarrow  dd\gamma, \quad   d \bar u \rightarrow d \bar u\gamma
\label{qqqq1}
\end{equation}
present a RAZ \cite{qqqqg} at
\begin{equation}
\cos\theta_\gamma = (Q_2 - Q_1)/(Q_2+Q_1),
\end{equation}
where $Q_1$ and $Q_2$ are the electric charges for initial state quarks
and $\theta_\gamma$ the photon scattering angle with respect to the
incident quark. But  some other processes such as
\begin{equation}
 u \bar u \rightarrow  u\bar u \gamma,  \quad  u  d \rightarrow u d\gamma,
\quad
 d \bar d \rightarrow  d \bar d\gamma, \quad   d u \rightarrow d u\gamma
\label{qqqq2}
\end{equation}
do not.  The cause for the difference is the ``same-sign rule'', as stated
earlier.
The locations of the zeros are clearly sensitive to the fractional charges
of the quarks \cite{qqqqg}, although there are no triple vector-boson
self-interactions
involved.  However, after convoluting with the hadron structure functions,
the RAZ becomes a dip. At low energies
where the valence quarks dominate, there is a good chance one could
find  the RAZ in these processes. Our experimental colleagues may
consider to re-examine the low energy  data, such as that at CERN ISR $pp$
collider  ($\sqrt s \sim$ 30 - 60 GeV), for this purpose.
At higher energies, such as at the Fermilab Tevatron \cite{eqeqg1},
the QCD multiple-jet  processes would completely swamp the RAZ signal.

It is straightforward to calculate the processes
$e^\pm p \rightarrow  e^\pm X \gamma$  \cite{eqeq,eqeq2}  by simply replacing
one of the quarks by $e^\pm$ in processes \ref{qqqq1},\ref{qqqq2}.
Once again, at low energies where the valence quarks dominate,
it is possible to examine the dip resulted from the RAZ. At  HERA energies,
however,  the RAZ effects in $e^-p$ collisions seem to be largely washed
out, while it is claimed to be more promising in $e^+p$ collisions
\cite{eqeq2},
again due to the argument of  the ``same-sign rule''.
Inclusion of more realistic experimental
simulation may further worsen this situation.

\section*{III. \, RAZ in Theories Beyond the Standard Model}

The RAZ is a general feature in gauge theories. There are in fact many
more processes beyond the SM  in which  the RAZ occur.
The RAZ  theorem has been generalized to supersymmetric theories with
massless gaugino emission \cite{susy1} and RAZ have been found in the
exact supersymmetric limit for processes \cite{susy2} such as
\begin{equation}
d\bar u \rightarrow \tilde{W} \tilde{ \gamma}, \quad
\gamma e \rightarrow \tilde{W} \tilde{ \nu_L^{}} \quad {\it etc.}.
\end{equation}
In this limit, the RAZ locate at the same places as those
for the SM partners.

The RAZ is also found in
charged Higgs boson production $p\bar p \rightarrow H^\pm \gamma$
\cite{higgs1}, although the small Yukawa coupling of  $H^\pm$ to
light fermions would make this process unobservable. A more promising
process is for the decay $H^+ \rightarrow t \bar b \gamma$
\cite{higgs2} if  kinematically accessible.
Similarly,  the RAZ effects in
radiative decays of other charged scalar particles such as
lepto-quarks are also studied \cite{lepto}.

\section*{IV. \, Summary}

Certain tree-level processes involving massless gauge bosons and charged
particles present  radiation amplitude zeros (RAZ).  With higher order
radiative corrections and in a more realistic experimental environment,
those zeros are always approximate  or become dips. In the SM
with a spontaneously broken gauge symmetry, $WZ$ final state
develops approximate zeros at high energies. In general,
the nature of  those zeros  is sensitive
to the gauge couplings of   vector bosons and that of fermions  as well.
Studying these RAZ experimentally
may thus provide probes to physics beyond the SM.

Progress has been made in studying the RAZ both theoretically
and experimentally, {\it e. g.}
$p \bar p \rightarrow W^\pm \gamma \rightarrow l^\pm \nu \gamma$
and  $W^\pm Z \rightarrow l^\pm \nu l^+ l^-$
at Fermilab Tevatron energies.  Other processes such as
 $qq' \rightarrow  qq' \gamma$ at low energies, and
$e^+ p \rightarrow  e^+ X \gamma$ at HERA
should be  examined at a level with  realistic experimental
acceptance to draw further conclusion.
It is clearly challenging to experimentally observe  those
``approximate'' zeros.  Hopefully one day,  we would be able
not only to observe the RAZ, but also in so doing to find
 some hints on new physics in the vector-boson sector.

\section*{Acknowledgement}

I would like to thank Uli Baur, Steve Errede and Thomas M\"uller
for their invitation and their nice organization of  such a  stimulating
conference. I am also grateful to Bob Brown for illuminating discussions
during the conference.  This work was supported in part by
the U.S.~Department of Energy
under Contract No.  DE-FG03-91ER40674.

%%%%%%%%%%%%%%%%%% figure captions %%%%%%%%%
%
\begin{figure}[t] % fig 1
\caption{\label{wgzero}
Differential cross sections
$d\sigma(\lambda_{\rm w},\lambda_{\rm \gamma})/d\cos\theta$
for  a). $e^- \bar \nu_e \rightarrow W^- \gamma$ and
b). $d\bar u \rightarrow W^- \gamma$, where $\theta$ is the
polar angle between $W^-$ and the incident fermion ($e^-$ or $d$)
in the c.m. frame.
For comparison, the differential  cross section for $e^+e^-\rightarrow ZZ$,
in which there is no RAZ,  has been included in  a).}
\end{figure}
\begin{figure}[t] % fig 2
\caption{\label{wzzerofig}
Differential cross sections
$d\sigma(\lambda_{\rm w},\lambda_{\rm z})/d\cos\theta$
for  a). $e^- \bar \nu_e \rightarrow W^- Z$ and
b). $d\bar u \rightarrow W^- Z$, where $\theta$ is the
polar angle between $W^-$ and the incident fermion ($e^-$ or $d$)
in the c.m. frame. For comparison, the differential  cross section for
$e^+e^-\rightarrow ZZ$, in which there is no RAZ,  has been included in  a).
}
\end{figure}
\begin{figure}[t] % fig 3
\caption{\label{hagifig}
Differential cross sections a). $d \sigma/d\cos\theta^*_+$
b). $d \sigma/d\cos\theta^*_-$ for $p \bar p \rightarrow W^- \gamma$,
$W^- \rightarrow e^- \bar \nu_e$.  Note that  $\theta^*$ here is the
polar angle between $\gamma$ and $p$ in the c.m. frame.
Effects from an anomalous coupling $\kappa$ are also shown,
where $\kappa=1$ corresponds to the SM results.
Acceptance cuts are described in Ref. \protect\cite{hagi1}. }
\end{figure}
\begin{figure}[t]  % fig.4
\caption{\label{yqcd}
The differential cross section for the photon rapidity in the
reconstructed center of mass frame for the reaction
$p \bar p \to W^+ \gamma \to e^+ \nu_e \gamma$ at
$\protect\sqrt{s} = 1.8$~TeV in the SM.
a) The inclusive NLO differential cross section (solid line) is
shown, together with the ${\cal O}(\alpha_s)$ 0-jet (dotted line)
and the (LO) 1-jet (dashed line) exclusive differential cross sections.
b) The NLO $W\gamma +0$~jet exclusive differential cross section
(dotted line) is compared with the Born-level LO
differential cross section (dot-dashed line).
A jet is defined as $p_T^j > 10$ GeV and $|\eta^j| < 2.5$.
Other cuts imposed are described in  Ref. \protect\cite{bhowg}. }
\end{figure}
\begin{figure}[t] % fig5
\caption{\label{zgwg}
The ratio of integrated cross sections
as a function of the minimum transverse momentum of the photon,
$p_T^{min}(\gamma)$, at the Tevatron. The dashed line shows the
corresponding ratio of $Zj$ to $W^\pm j$ cross sections for comparison.
 }
\end{figure}
\begin{figure}[t]  % fig.6
\caption{\label{2dcorrl}
The double differential distribution $d^2\sigma/dy_\gamma dy^{}_W$ for
$p\bar p\rightarrow W^+\gamma \rightarrow \ell^+ \nu \gamma$,
$\ell=e,\,\mu$, in the Born approximation at the Tevatron
(1.8~TeV). The cuts imposed are described in
Ref. \protect\cite{ycorrl}.
}
\end{figure}
\begin{figure}[t]  % fig.7
\caption{\label{1dcorrl}
The pseudorapidity difference distribution,
$d\sigma/d\Delta\eta(\gamma,\ell)$, for $p\bar p\rightarrow W^+\gamma$,
$W^\pm  \rightarrow\ell^+ \nu$ with $\ell=e,\,\mu$, at  the Tevatron in the
Born
approximation for anomalous $WW\gamma$ couplings. The curves are for
the SM (solid), $\Delta\kappa_0=2.6$
(dashed), and $\lambda_0=1.7$ (dotted). Only one coupling is varied at a
time.  The error bars indicate the
expected statistical uncertainties for an integrated luminosity of
22~pb$^{-1}$.
The cuts imposed are described in   Ref. \protect\cite{ycorrl}.
}
\end{figure}
\begin{figure}[t]      % fig.8
\caption{\label{wzcorrl}
The differential cross section for the
rapidity difference $\Delta y(Z,\ell_1)$ for
$\protect{p\bar p \to}$ $\protect{ W^+Z + X \to\ell_1^+\nu_1\ell_2^+
\ell_2^- + X}$  at $\protect{\sqrt{s} = 1.8}$~TeV.
The solid and dot-dashed curves show the inclusive NLO and the
LO SM prediction, respectively. The dashed and dotted lines give the
results for $\Delta\kappa^0=+1$ and $\Delta\kappa^0=-1$. The error bars
associated with the solid curves indicate the expected statistical
uncertainties for an integrated luminosity of 10~fb$^{-1}$. The cuts
imposed are  described in Ref. \protect\cite{bhowz}.}
\end{figure}

\end{document}